\documentclass[12pt,preprint]{aastex}
\def\beq{\begin{equation}}
\def\eeq{\end{equation}}
\def\bey{\begin{eqnarray}}
\def\eey{\end{eqnarray}}
\def\event{sc26\_2218}

\def\kms{{\rm \,km\,s^{-1}}}

\def\kpc{\,{\rm {kpc}}}

\def\chisq{\chi^2}

\def\tE{t_{\rm E}}
\def\rEt{\tilde{r}_{\rm E}}

\def\u0{u_0}

\def\mI0{m_{I,0}}
\def\ustar{\rho_\star}
\def\thetastar{\theta_\star}
\def\thetaE{\theta_{\rm E}}
\def\fs{f_{\rm s}}

\def\m{~\rm{mag}}

\def\au{{\rm AU}}
\def\mas{{\rm mas}}
\def\mias{{\rm $\mu$as}}

\newcommand{\zdot}{\makebox[0pt][l]{.}}
\newcommand{\up}[1]{\ifmmode^{\rm #1}\else$^{\rm #1}$\fi}

\newcommand{\arcd}{\ifmmode^{\circ}\else$^{\circ}$\fi}
\newcommand{\arcm}{\ifmmode{'}\else$'$\fi}
\newcommand{\arcs}{\ifmmode{''}\else$''$\fi}

\begin{document}

\title{A Brown Dwarf Microlens Candidate in the OGLE-II Database}


\author{Martin C. Smith\footnote{Univ. of Manchester, Jodrell Bank
    Observatory, Macclesfield, Cheshire SK11 9DL, UK;
    msmith@jb.man.ac.uk, smao@jb.man.ac.uk}, Shude Mao\footnotemark[1], and
    Przemys{\l}aw Wo\'zniak\footnote{Los Alamos National Laboratory,
    MS D436, Los Alamos, NM 87545, USA; wozniak@lanl.gov}}

\begin{abstract}
We describe a unique mass determination for a microlensing event
from the second phase of the Optical Gravitational Microlensing
Experiment (OGLE-II). The event, \event, which is very bright
(baseline magnitude $I=15.10$), appears to 
exhibit both parallax and finite source effects. The parallax
effect allows us to determine the projected Einstein radius
on the observer plane ($\rEt \approx 3.8 \au$), while the
finite source effect allows us to determine the ratio of
the angular source size and the angular Einstein radius. As the
angular size of the star can be estimated using its color and
magnitude, we can hence determine the angular Einstein radius
$\thetaE \approx 0.1 \mas$. By combining $\rEt$ and $\thetaE$ we can
determine the lens mass $M \approx 0.050^{+0.016}_{-0.011} M_\odot$,
independent of the source distance. The lens is therefore 
a brown dwarf candidate. However, the ``parallax'' signature
is weak and so we cannot completely discount the possibility that these
signatures originate from binary rotation of the source (which would
prevent any estimate of the lens mass), rather than parallax. However,
this can be tested by future spectroscopic observations.
This event highlights the
scientific returns for intense monitoring of bright microlensing
events, since the parallax and finite source effects can be more
easily identified due to their high signal-to-noise ratios.
\end{abstract}


\section{Introduction}

Gravitational microlensing events toward the Galactic bulge provide
a unique sample to study the mass functions in the Milky Way
(for review on microlensing, see  Paczy\'nski 1996). However, this
important application is severely hampered by the
lens degeneracy. This degeneracy arises because, in general, only
a single physical constraint can be derived from an observed light 
curve, namely the Einstein radius crossing time, which depends on the
lens mass, various distance measures and the relative velocity. This
degeneracy can be most easily lifted using space satellites (e.g.
Gould 1994b), such as the proposed GEST (Bennett et al. 2002a). 
Without observations from space, unique mass determinations are
only possible for exotic microlensing events. So far, this
has been possible for no more than two microlensing events. An et
al. (2002) determined the lens mass for the binary caustic crossing
event, EROS BLG-2000-5, by
combining parallax effects (Gould 1992) and finite source effects
(Gould 1994a; Witt \& Mao 1994; Nemiroff \& Wickramasinghe 1994).
Alcock et al. (2001) have also made a tentative estimate of the
microlens mass for a different event, MACHO-LMC-5, by utilizing
measurements of both the parallax effect and the microlens proper
motion. However, this mass determination requires confirmation since
it relies on a difficult measurement of the very small microlensing
parallax effect.

In this paper we present another unique mass determination for event
\event\, from OGLE-II Difference Image Analysis data. 
Using high-precision data we are able to detect both parallax and
finite source effects, which allows us to determine the mass of
the lensing object.

\section{Observational Data}

The observational data that we use are based on observations performed
in the second phase of the OGLE experiment (OGLE-II). The observations
were carried out with the 1.3-m Warsaw 
telescope at the Las Campanas Observatory, Chile, which is 
operated by the Carnegie Institution of  Washington. The 
instrumentation of the telescope and CCD camera are described in detail
by Udalski, Kubiak \& Szyma{\'n}ski (1997). 

The event \event, which was observed toward the Galactic bulge,
was first identified in Wo\'zniak et al. (2001) using
the difference image analysis software of Wo\'zniak (2000). 
The light curve data for this event are available 
publicly\footnote{\tt http://astro.princeton.edu/\~\,wozniak/dia/lens/}.
The coordinates of the star are
${\rm RA}=17:47:23.29$, ${\rm DEC}=-34:59:52.4$ (in J2000), which
correspond to ecliptic coordinates,
$\lambda=267\zdot\arcd364$, $\beta=-11\zdot\arcd 587$ (in J2000), and
Galactic coordinates $l=355\zdot\arcd0084$, $b=-3\zdot\arcd4640$.

We extracted the I-band baseline magnitude of the star from
Wo\'zniak et al. (2001) and cross-checked with the OGLE-II DoPhot
Photometry. The star turns out to be very bright with a baseline
magnitude of $I_{\rm base}=15.10 \pm 0.02$ mag. There are 8 $V$-band
observations for the star.
We used the three most reliable of these to derive the $V$-band
magnitude, which gives the color of the star
$V-I=1.76\pm 0.03$ mag.\footnote{
Udalski et al. (2002) recently published the 
the photometry and BVI colors for 30 million stars toward
the Galactic center. However their values for this star cannot be used,
as both its baseline magnitude and colors are
based on average values that include the microlensed data points.}

\begin{figure}
\plotone{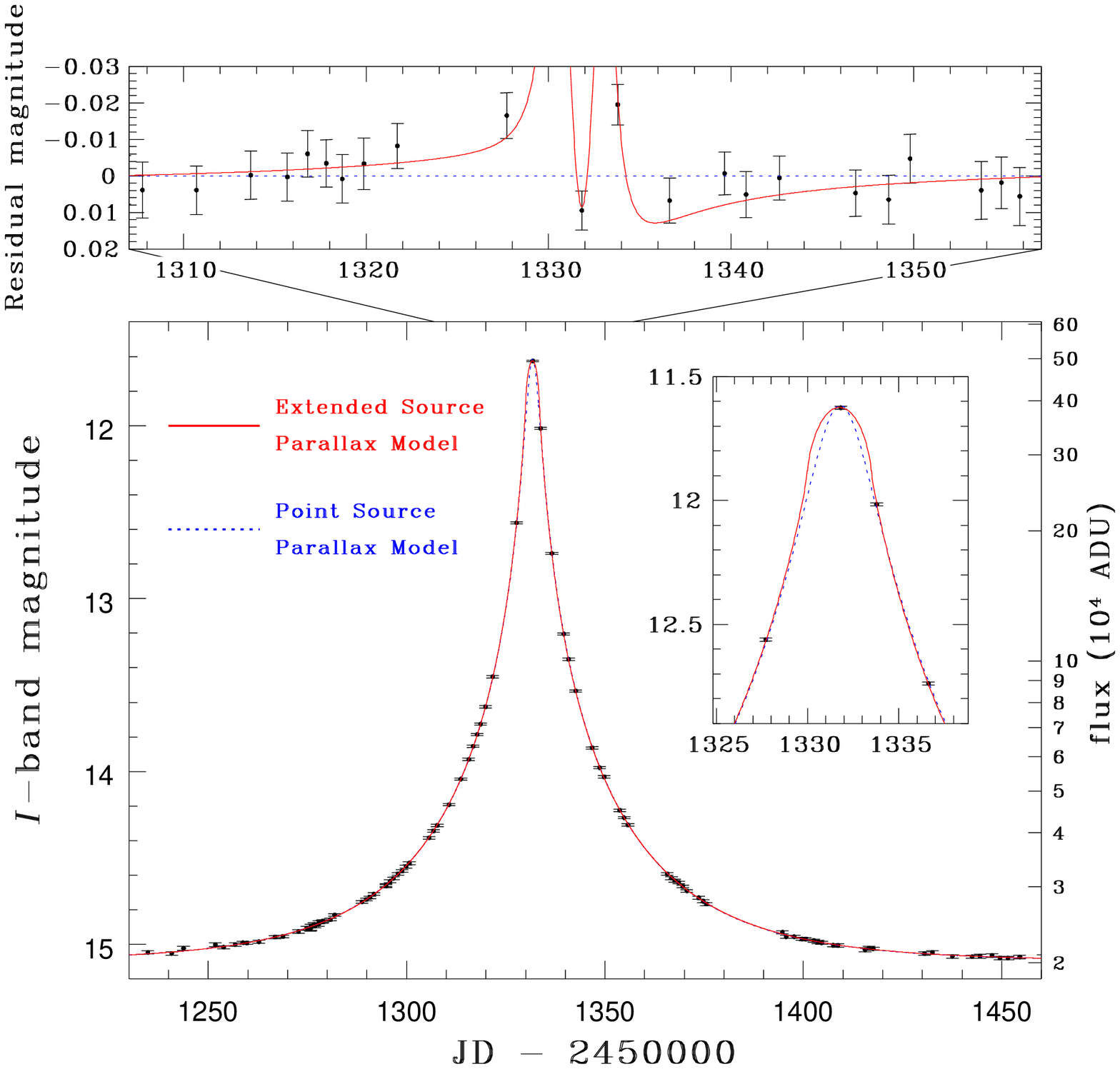}
\caption{The lower panel shows the
I-band light curve for the microlensing event \event. The magnitude
scale is shown on the left $y$-axis, while the flux, which is units of
$10^4$ ADU, is shown on the right (logarithmic) 
$y$-axis. The dotted line is the point-source parallax model while the
solid line is the model that accounts for both parallax and 
finite source effects. The top panel shows the residual magnitude from
the parallax model (i.e. the observed data points subtracted by the
parallax model), while the solid line shows the prediction for
the best model that accounts for both parallax and finite
source size effects. This plot shows only the microlensed region of
the light curve -- the full data-set consists of 3 seasons
($550~{\rm days} \la t = {\rm JD} - 2450000 \la 1500~{\rm days}$).
\label{fig:lc}
}
\end{figure}

Fig. \ref{fig:cmd} shows the color-magnitude 
diagram for the field around \event. The star
appears to lie within the red clump giant region and therefore the source
is probably located in the Galactic bulge. The light curve obtained
from the difference image analysis is shown in Fig. \ref{fig:lc}. 
The quality of the light curve is striking, with a typical
error of 0.8\% throughout a large magnitude range ($I=15.1$ to 11.6 mag).

\section{Model}

\subsection{Point Source models}

We first fit the observed light curve with the usual standard
and parallax microlensing models. Both models use the approximation
that the source can be regarded as point-like. The fitting procedures
are identical to those used in our previous papers (Smith et al. 2002b;
Mao et al. 2002). We refer the readers to those papers for details;
here we only briefly review the parameters that we use in the models.

To fit the {\it I}-band data with the standard model incorporating
blending (see, e.g. Paczy\'nski 1986), we need five parameters, namely
the impact parameter $\u0$, the time of closest approach
$t_0$, the baseline magnitude $I_{\rm base}$, the Einstein-radius
crossing timescale, $\tE$, and the blending parameter, $\fs$,
which parametrizes the fraction of light that
is contributed by the lensed source. The parameter ($\fs$)
is required to account for blending, owing to the crowded
nature of the fields toward the Galactic center.

The best-fit parameters are found by minimizing the usual 
$\chisq$ using the MINUIT program in the
CERN library$\footnote{http://wwwinfo.cern.ch/asd/cernlib/}$
and are tabulated in Table 1. The resulting $\chisq$ for the standard
model is 277.9 for 236 data points.

Next we proceed to incorporate the parallax effect (Gould 1992). This
effect, which arises due to the Earth's orbital motion, produces
perturbations from the standard model light curve, and is described in
detail in Soszy\'nski et al. (2001); see also Alcock et al. (1995),
Dominik (1998). It requires two additional parameters;
the Einstein
radius projected onto the observer plane, $\rEt$, and an angle in the
ecliptic plane, $\psi$, describing the orientation of the lens
trajectory (given by the angle between the heliocentric ecliptic 
$x$-axis and the normal to the trajectory)

The best-fit parallax model is given in Table 1, and the corresponding
light curve is presented in Fig. \ref{fig:lc}. This model clearly
improves upon the standard model, with a best-fit parallax $\chi^2$ of
236.7. It is interesting to note that one would not normally
expect to observe parallactic deviations for an event with such a
short time-scale ($\tE = 39.2$ days for this point-source parallax
model), since the Earth has not moved a significant distance in its
orbit around the Sun. However, due to the high signal-to-noise ratio
for \event, we are able to detect these weak parallactic
signatures. Also, as a result of the weak signatures, a slightly worse 
alternative parallax model is possible (see \S4).

\subsection{Finite Source Models}

Notice that the parallax model does not match the observed
fluxes around the peak when the source is highly
magnified (the peak magnification for \event\, is approximately
25). This kind of deviation at high magnifications is precisely
what is expected from finite source effects (Gould 1994a; Witt \&
Mao 1994; Nemiroff \& Wickramasinghe 1994). We therefore proceed to
incorporate this effect into our model. This requires one additional
parameter, $\rho_\star$, which denotes the source radius in units of
the lens' angular Einstein radius.

We initially attempt to fit the light curve with a finite source model
without incorporating the parallax effect. This is clearly unable
to provide a suitable fit, with $\chi^2=261.3$. The next logical step is to
incorporate the parallax effect and the finite source effect
simultaneously in the modeling. This provides a greatly improved fit, with
$\chi^2=214.4$, i.e. an improvement of $\Delta \chi^2=22.3$ on the
next best model (the point-source parallax model). The significance of
this improvement is $4.7\sigma$. We also attempted to fit the light
curve with a non-uniform source brightness, using the limb-darkening
profile of An. et al. (2002), but this gave no noticeable improvement
and the best-fit parameters remained unchanged.
 
The best-fit parameters for this finite source parallax model are
given in Table 1 and the corresponding light curve is presented in
Fig. \ref{fig:lc}. The most important parameters from this model are,
\beq
\rEt=3.8^{+1.2}_{-0.8}{\rm \, AU},~~\tE=39.5\pm1.0{\rm \, days},~~\rho_\star=0.0659\pm0.0005.
\label{rho}
\eeq
Unfortunately the value for $\rEt$ is not well constrained, but
$\rho_\star$ is tightly constrained.
From these quantities we can calculate
$\tilde{v}=\tilde{r}_E/t_E=167^{+53}_{-35} \kms$,
the transverse velocity of the lens relative to the source, projected
onto the observer plane. This is relatively high compared to previous
reported parallax events (cf. Smith, Mao \& Wo\'zniak 2002a; Bennett
et al. 2002b), although this could be due to the fact that the
timescale is unusually short for a microlensing parallax event and
also because we expect the projection factor to be large (since the
lens is predicted to reside much closer to the source than the observer;
see \S \ref{mass}).

\subsection{Mass determination}
\label{mass}

\begin{figure}
\plotone{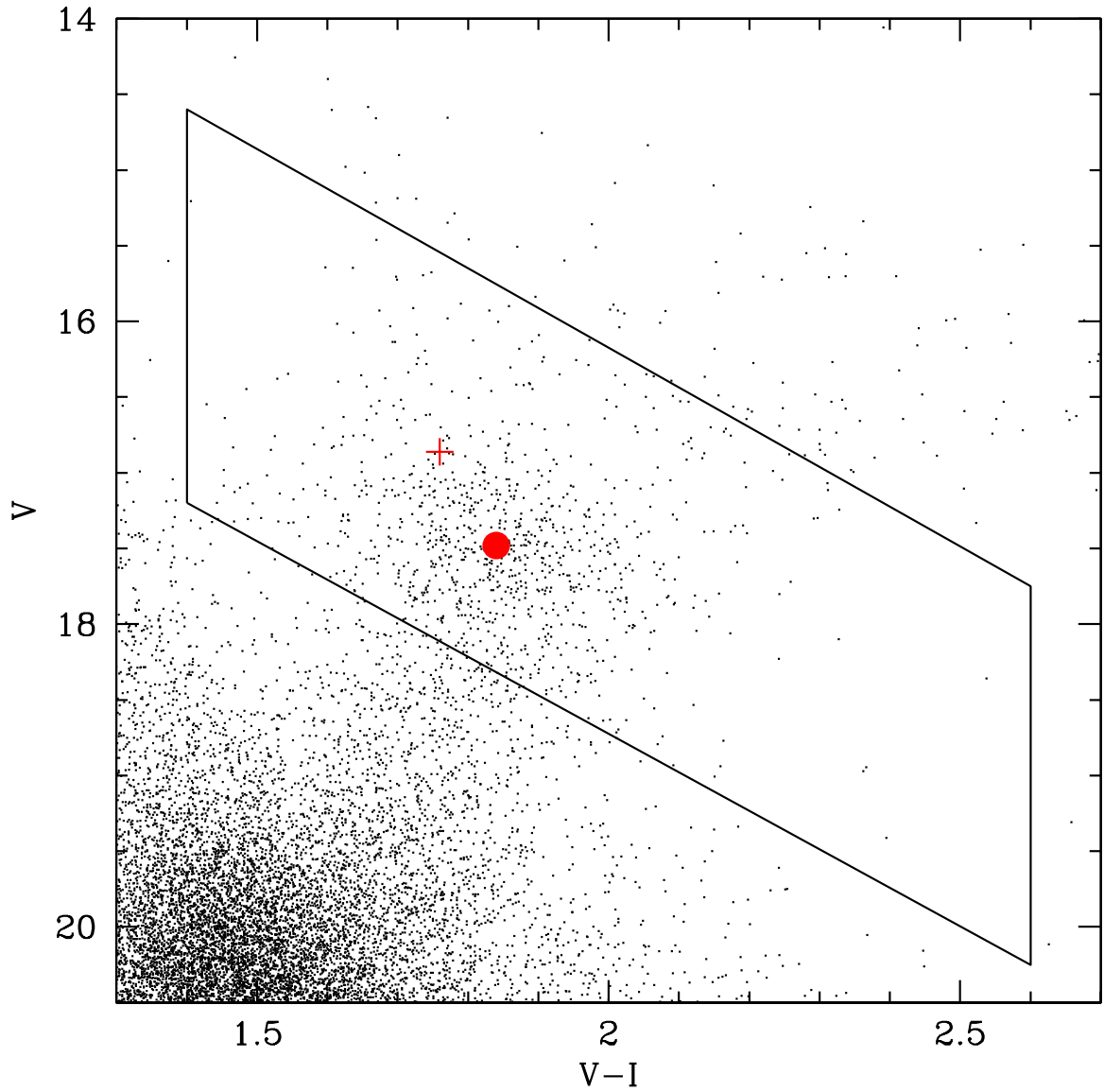}
\caption{The color-magnitude diagram for the $5' \times 5'$ field
around \event. The position of the lensed star is denoted by a plus sign
(+), while the center of the red clump giant region is denoted by a
filled circle ($\bullet$). The star lies within the
red clump region and is therefore probably located in the Galactic
bulge. The box indicates where the red clump giants are selected
(cf. Wo\'zniak \& Stanek 1996).
\label{fig:cmd}
}

\end{figure}

Given the above model incorporating both parallax and finite source
effects, it is possible to determine the lens mass provided that an
estimate of the angular size of the source can be made.
Albrow et al. (2000) showed that  the angular size of the source can
be determined from its de-reddened color and magnitude. The
color-magnitude diagram for this event (Fig. \ref{fig:cmd}), shows
that the observed magnitude of the center of the red clump region for
this field is given by 
$I_{\rm{cl, obs}} \approx 15.62~\rm{mag}$, $(V-I)_{\rm{cl, obs}}
\approx 1.85~\rm{mag}$. From Popowski (2000), the intrinsic
de-reddened color of the center of the red clump region in the
Galactic center is $(V-I)_{\rm{cl, 0}} = 1.00~\rm{mag}$. Therefore
the reddening for the field is given by,
\beq \label{red}
E(V-I)_{\rm{cl}} = (V-I)_{\rm{cl, obs}} - (V-I)_{\rm{cl, 0}} \approx
0.85~\rm{mag}.
\eeq
As the extinction in the Galactic center may be anomalous
(Udalski 2002), we checked the extinction slope using the method
of Wo\'zniak \& Stanek (1996). We find that
the extinction slope, $\Delta I/\Delta(V-I)$,
is about 1.39, which is close to, but slightly smaller
than, the standard value (1.49, Stanek 1996).
Therefore the extinction is,
\beq \label{ext}
A_{I,\rm{cl}}=1.39 \times E(V-I)_{\rm{cl}} \approx 1.18\m.
\eeq
Since our best-fit model favors no blending, the
$I$-band baseline apparent magnitude of the lensed source is $I_{\rm s}=15.10\m$.
As was stated in \S 2, from the color-magnitude diagram for this
field (Fig. \ref{fig:cmd}) one concludes that the source is 
likely to be located in the red clump region and hence undergo the
same reddening and extinction as calculated in eq. (\ref{red}) \&
(\ref{ext}). This would imply that the source has an intrinsic
brightness of $I_{\rm s, 0} \approx 13.92\m$ and an intrinsic color of
$(V-I)_{\rm s, 0} \approx 0.91\m$.

This intrinsic color can then be converted from $(V-I)_{\rm s, 0} \approx
0.91\m$ into $(V-K)_{\rm s, 0}\approx 2.05\m$ (using, for example, Table III
of Bessell \& Brett 1988). Once this has been done, a value for the
angular size of the source ($\theta_\star$) can be calculated from
an empirical surface brightness-color relation (eq. 4 from Albrow et
al. 2000), and we find that $\thetastar=7.07$ \mias, which corresponds to a
physical stellar radius of about $12R_\odot$ if the source is $8\kpc$ away.

This value for $\theta_\star$ can be combined with the above
constraint on $\rho_\star$ (eq. \ref{rho}) 
to determine the lens' angular Einstein radius,
\beq
\label{eq:thetaE}
\theta_E=\frac{\theta_\ast}{\rho_\ast} = 107.35 \pm 7.52 ~\mu\rm{as}.
\eeq
Combining $\thetaE$ and $\rEt$, we can infer the lens mass
directly,
\beq
M = \frac{c^2}{4G}\rEt\thetaE 
= 0.050^{+0.016}_{-0.011} M_\odot.
\eeq
The errors on the lens mass are dominated by the errors in $\rEt$
(see eq. \ref{rho}), since the rest of the parameters are
comparatively well constrained compared to $\rEt$.
It is important to stress that the lens mass is independent of
the source distance.\footnote{It should be noted that the estimate of
$\rho_\star$ is not entirely independent of the source distance, because
if the source lies significantly closer to us than the Galactic center
our assertion that it undergoes the same amount of
extinction/reddening as the red clump stars may be invalidated.}
However, the distance to the lens depends on the
source distance; if the source is a red clump star and it is located
in the bulge at a distance of $8\,\rm{kpc}$, then this implies that
the distance to the lens is approximately $6.5\,\rm{\rm kpc}$.
At such a distance, the brown dwarf lens should contribute essentially no
light, which is entirely consistent with the best-fit blending
parameter $\fs=1$.

\section{Discussion}

In this paper we have presented a detailed analysis 
of the microlensing event \event, identified first
by Wo\'zniak et al. (2001) using the OGLE-II database.
We showed that this event is best fitted by incorporating both 
parallax and finite source effects. 
Combined with an estimate of the source angular
size using its color and magnitude, these two
effects allow us to determine the lens mass uniquely.
Intriguingly, the lens mass ($M \approx
0.050^{+0.016}_{-0.011}M_\odot$) lies within the
brown dwarf regime. In comparison, An et al. (2002)
determined a total lens mass of $M \approx 0.61M_\odot$ and mass ratio
$q \approx 0.75$ for the binary lens EROS BLG-2000-5, and Alcock et
al. (2001) made a tentative estimate of $M \sim (0.04 - 0.1) M_\odot$
for MACHO-LMC-5.

The event was not identified in our first 
search (Smith et al. 2002a) because
at the time no baseline flux information was
available to us. As a result, the best-fit standard
model for this event was comparable to the best-fit parallax 
model. However, if we take into account the positive
definite constraint on the source flux the
standard model becomes worse than the parallax model. This
prompted us to pursue \event\, further in this study.
We have since redone the fitting for all events in the OGLE-II
catalog. The results for most events are unchanged, except
for \event\, and an additional parallax event (sc5\_2859) that
was reported in Smith (2003). This brings the total 
number of convincing parallax events in the OGLE-II catalog to
5/520, i.e. around 1\%.

It has been shown (Smith, Mao \& Paczy\'nski 2002c) that under certain
circumstances seemingly parallactic light curves can be well fit by a
model incorporating a constant acceleration term. This is important
for events that exhibit only weak perturbations from the standard
model, such as \event. Indeed, we found that a constant-acceleration
finite source model is able to provide a good fit to this event,
with $\chi^2=217.1$. This fit is only slightly worse than the best-fit
finite source parallax model ($\chi^2=214.1$).
For perturbations which are well described by a constant acceleration,
it is difficult to determine whether they originate from the 
parallax effect or from binary rotation in the source (Smith et
al. 2002c). Hence we cannot categorically state that these are
indeed parallactic signatures. However, this binary source hypothesis
can be tested through future spectroscopic observations, which should
reveal periodic modulations in radial velocity from the binary
rotations. The radial velocity is expected to be about $30{\rm
km\,s^{-1}}$ with a period of $\sim 1$ year. As the star is bright
($I\approx 15.10$ and $V\approx 16.86$), these observations should be
straightforward to perform even with medium-size telescopes.

Smith et al. (2002c) also showed that events which are suitably fit by
a constant-acceleration model exhibit two degenerate sets of parallax
solutions. For \event\, two sets of finite-source-size parallax
solutions were found with $\Delta \chi^2=0.13$. However, the important
parameters that determine the lens mass (i.e. $\rEt$ and $\rho_\star$)
are virtually identical between the two solutions, and the above
conclusions regarding the lens mass are consequently unaffected by
this degeneracy.

Assuming that this best-fit parallax model is accurate, it may be
possible to indirectly determine the proper motion of the lens. To do
this we require the transverse proper motion of the lens relative to
the source, which is directly determined from the parallax fit (but
not in the constant acceleration model). Given this
relative proper motion, if the absolute proper motion of the source
can be found using many background quasars in the field (see, for
example, Sumi, Eyer \& Wo\'zniak 2002) this will provide a
determination for the absolute proper motion of the lens. This
information can be used to provide further constraints on the lens for
any future likelihood analysis.

\event\, highlights the possibility of unique mass
determinations using highly-magnified bright microlensing
events. For such events, the signal-to-noise ratio
may be high enough that even minute parallax signatures
can be detected. If the sampling at the peak of the
light curve is dense enough, then finite source
effect may also be detected with high significance.
If the sampling for \event\, was higher, then
the finite source effect would have been detected at
even greater significance. With an order
of magnitude increase in the microlensing rate
for OGLE-III, surely many events similar to \event\, can
be discovered.

\acknowledgments{
We thank Bohdan Paczy\'nski for discussions and encouragements
on the manuscript, and the OGLE collaboration, in particular Andrzej
Udalski and Igor Soszy\'nski for access to and analysis of their data.
This work was partly supported by funds from proposal \#09518 provided
by NASA through a grant from the STScI, which is operated by the
AURA, under NASA contract NAS5-26555. MCS acknowledges the financial
support of a PPARC studentship.}


\clearpage

\begin{table}
\begin{center}
\tiny
\caption{The best standard model (first row), the best
parallax model (second row) and the best parallax
model incorporating finite source effects (third row) for
\event.}
\vspace{0.3cm}
\begin{tabular}{cccccccccc}
\tableline\tableline
 & $t_0$ & $\tE$ (day) & $\u0$ & $I_{\rm base}$
& $\psi$ (radians) & $\rEt$ (AU) & $\fs$ & $\ustar$ & $\chisq$  \\
\tableline
S & $1331.81 \pm 0.01$ & $45.31\pm 0.07$ & $0.0404\pm 0.0002$ &
$15.0991\pm 0.0007$
& --- & --- & $1.0000^{+ 0}_{-0.0006}$ & --- & 277.9/231 \\
P & $1330.0^{+0.7}_{-0.6}$ & $39.16^{+0.99}_{-0.96}$ & $0.1077\pm 0.0025$
 & $15.0974\pm 0.0007$ & $0.241^{+0.024}_{-0.018}$ &
 $3.17^{+0.93}_{-0.55}$ & $1.0000^{+ 0}_{-0.0023}$ & --- & 236.7/229 \\
P+F & $1331.1^{+1.0}_{-0.8}$ & $39.53^{+0.97}_{-0.96}$ & $0.064^{+0.034}_{-0.036}$
 &  $15.0975 \pm 0.0007$ & $0.285^{+0.060}_{-0.031}$ & $3.8^{+1.2}_{-0.8}$ &
 $1.0000^{+ 0}_{-0.0029}$ & $0.0659^{+0.0005}_{-0.0005}$ & 214.4/228 \\ 
\tableline
\end{tabular}
\end{center}
\end{table}
\normalsize 

\clearpage



\begin{references}

\reference{} Albrow, M. D., et al. 2000, \apj, 534, 894 
\reference{} Alcock, C., et al. 1995, \apj, 454, L125
\reference{} Alcock, C., et al. 2001, Nature, 414, 617
\reference{} An, J. H., et al. 2002, \apj, 572, 521
\reference{} Bessell, M. S., \& Brett, J. M. 1988, PASP, 100, 1134
\reference{} Bennett, D. P., et al. 2002a, astro-ph/0209435
\reference{} Bennett D.P., et al. 2002b, ApJ, 579, 639
\reference{} Dominik, M. 1998, \aap, 329, 361
\reference{} Gould, A. 1992, \apj, 392, 442
\reference{} Gould, A. 1994a, \apj, 421, L71
\reference{} Gould, A. 1994b, \apj, 421, L75
\reference{} Mao, S. et al. (OGLE collaboration) 2002, \mnras, 329, 349
\reference{} Nemiroff, R. J., \& Wickramasinghe, W. A. D. T., 1994, ApJ, 424, L21
\reference{} Paczy\'nski, B. 1986, \apj, 304, 1
\reference{} Paczy\'nski, B. 1996, \araa, 34, 419
\reference{} Popowski, P. 2000, ApJ, 528, L9
\reference{} Smith, M. C., Mao, S., \& Wo\'zniak, P. 2002a, \mnras, 332, 962
\reference{} Smith, M. C., et al. (OGLE collaboration) 2002b, \mnras, 336, 670
\reference{} Smith, M. C., Mao, S., \& Paczy\'nski, B. 2002c, \mnras, in press (astro-ph/0210370)
\reference{} Smith, M. C. 2003, \mnras, submitted
\reference{} Soszy\'nski, I., et al. 2001 (OGLE collaboration), ApJ, 552, 731
\reference{} Stanek, K. Z. 1996, ApJ, 460, L37
\reference{} Sumi, T., Eyer, L., \& Wo\'zniak, P. R., astro-ph/0210381 
\reference{} Udalski, A., Kubiak, M., \& Szyma\'nski, M., 1997, Acta Astron.,
47, 319
\reference{} Udalski, A., et al. 2002, astro-ph/0210278
\reference{} Udalski, A. 2002, astro-ph/0210367
\reference{} Witt, H. J., \& Mao, S. 1994, ApJ, 430, 505
\reference{} Wo\'zniak, P. 2000, Acta Astron., 50, 421
\reference{} Wo\'zniak, P. \& Stanek, K.S. 1996, ApJ, 464, 233
\reference{} Wo\'zniak, P., et al. (OGLE collaboration)	2001, Acta Astron., 51, 175

\end{references}
\end{document}